\documentclass[epj]{webofc}
\usepackage[varg]{txfonts}   
\usepackage{graphics}
\usepackage{floatflt}

\woctitle{XLVI International Symposium on Multiparticle Dynamics}
\begin{document}
\title{Interactions of Cosmic Rays around the Universe}
%
%
\subtitle{Models for UHECR data interpretation}
\author{Denise Boncioli\inst{1}\fnsep\thanks{\email{denise.boncioli@desy.de}}
}

\institute{DESY, Platanenalle 6, D-15738 Zeuthen, Germany
          }

\abstract{%
Ultra high energy cosmic rays (UHECRs) are expected to be accelerated in astrophysical sources and to travel through extragalactic space before hitting the Earth atmosphere. They interact both with the environment in the source and with the intergalactic photon fields they encounter, causing different processes at various scales depending on the photon energy in the nucleus rest frame. UHECR interactions are sensitive to uncertainties in the extragalactic background spectrum and in the photo-disintegration models.}
%
\maketitle

\section{Introduction}
\label{intro}
The origin and nature of ultra high energy cosmic rays (UHECRs) are unknown: where these particles, as energetic as $\approx 10^{20}$ eV, come from, their chemical composition and which mechanisms are involved in their acceleration are still open issues. Extragalactic sources are considered as candidate for cosmic ray acceleration at the highest energies because of their characteristic size and magnetic fields \cite{Hillas:1985is}.\\
In order to predict the expected flux of UHECRs at Earth, interactions of particles with the photon fields both in the source and in the extragalactic space have to be taken into account. The suppression of the flux \cite{Valino:2015,Ivanov:2015} measured by UHECR experiments as the Pierre Auger Observatory \cite{ThePierreAuger:2015rma} and the Telescope Array \cite{AbuZayyad:2012kk} can be attributed to the energy losses suffered in these interactions and/or to the maximal energy reached at the source. If due to interactions, the suppression can be generated by different phenomena depending on the mass of the cosmic ray nucleus.\\
Measurements of observables sensitive to UHECR mass composition are not conclusive. The Telescope Array experiment claims to find a consistency of their $X_{\mathrm{max}}$ distributions with light composition above $10^{18}$ eV \cite{Belz:2015}. On the contrary, the Auger collaboration finds that the primary mass is decreasing reaching the minimal values at around $10^{18.3}$ eV and starts to increase for the higher energies; at the same time the spread of the mass is almost constant till $10^{18.3}$ eV and then starts to decrease, indicating that the relative fraction of protons becomes smaller for higher energies \cite{Porcelli:2015}. By comparing the $X_{\mathrm{max}}$ of the two experiments taking into account different resolutions, acceptances and analysis strategies, the two results are found to be in good agreement within systematic uncertainties \cite{Unger:2015}.\\
At the highest energies, the spectrum features are related to characteristics of propagation of cosmic rays in the extragalactic background fields.
If UHECRs above $10^{18}$ eV are mainly extragalactic protons, they interact with the cosmic microwave background (CMB) and the infrared, optical and ultraviolet background (named as extragalactic background light, EBL). The ankle, {\it i.e.} the hardening at about $5\times 10^{18}$ eV, in the proton dip model, is expected to be due to energy losses due to electron-positron pair production mainly on CMB \cite{Berezinsky:2002nc}. In the same context the suppression of the flux can be justified with the energy losses suffered by protons in photo-meson reactions, as it was predicted in \cite{Greisen:1966jv,Zatsepin:1966jv} (the GZK effect). The astrophysical scenario is complicated by the fact that source properties, as for example their distribution and their acceleration power, are unknown: the suppression of the flux, even in the pure-proton composition case, cannot be addressed to GZK cutoff in absence of knowledge about sources \cite{Aloisio:2010wv}.
Since the threshold energy for photo-meson production (roughly) increases $\propto A$, the suppression of the flux in case of a mixed composition is expected to be due, at least partly, to the photo-disintegration of nuclei, occurring after the excitation of the giant dipole resonance (GDR, $\approx$8 MeV in the NRF) that follows the absorption of a CMB or EBL photon \cite{Aloisio:2008pp,Aloisio:2010he}.\\
From the point of view of the models, a self-consistent interpretation of the spectrum and composition results of the Auger Observatory and the Telescope Array experiment can be used to limit the astrophysical scenarios compatible with data.

\section{Hypothesis of pure-proton composition}

\begin{figure*}
\centering
\includegraphics[width=6cm]{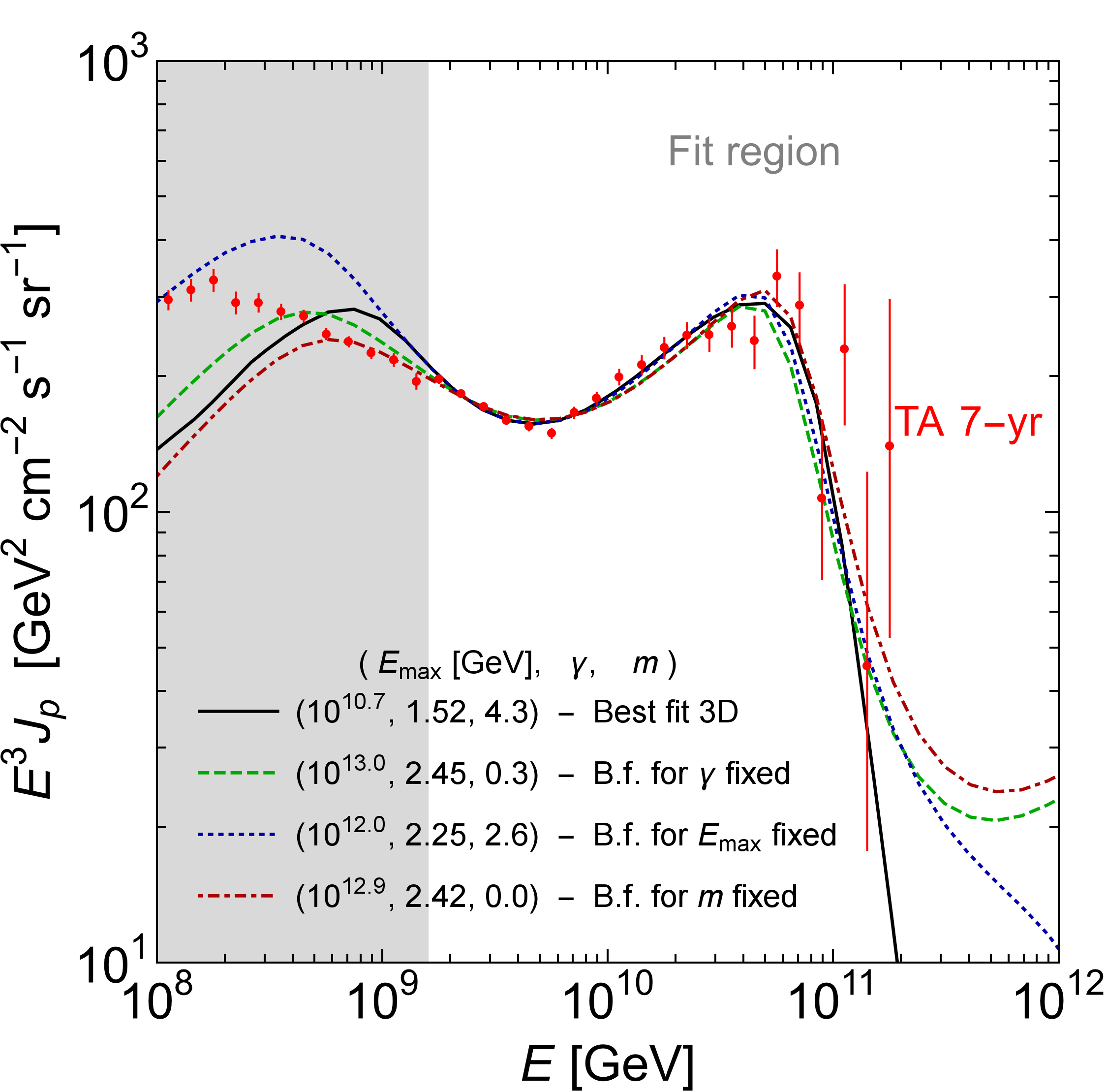}
\includegraphics[width=6cm]{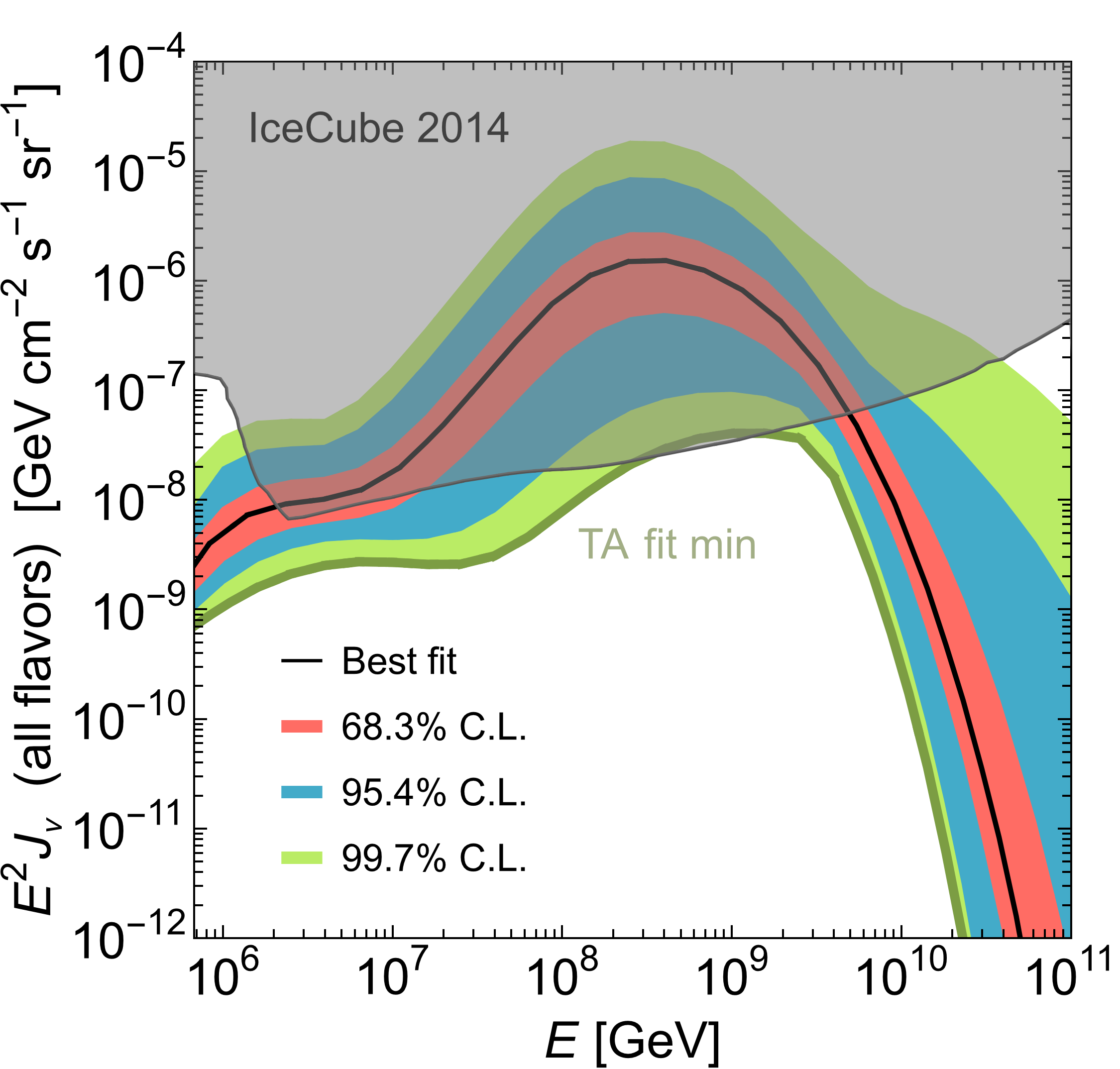}
\caption{Right: Best-fit spectrum (black solid line) superimposed to the TA energy spectrum \cite{Ivanov:2015}, as taken from \cite{Heinze:2015hhp}. Left: All-flavor flux of cosmogenic neutrinos predicted by the best fit to the TA energy spectrum \cite{Heinze:2015hhp}.}
\label{fig-1}     
\end{figure*}

The fundamental assumption in this case is that of an extragalactic pure proton UHECR composition, as in the proton dip model. This assumption can be tested in light of the UHECR energy spectrum measurements from the TA collaboration \cite{Ivanov:2015} and the upper bound on the flux of cosmogenic neutrinos from the IceCube collaboration \cite{Aartsen:2016ngq}, as done in \cite{Heinze:2015hhp}. The distribution of (identical) sources is taken as homogeneous, and the source evolution is parametrized relative to the star formation rate (SFR).
A fit of the TA spectrum is performed (and it is shown in Figure~\ref{fig-1}, left panel): the fit slightly prefers low maximal energies, hard
spectra, and strong source evolution. The high value of $m$
implies that the contribution of distant sources is enhanced with
respect to SFR evolution.\\ 
Cosmogenic neutrinos are produced by the decay of pions,
muons, kaons, and neutrons produced in photo-hadronic
interactions during cosmic-ray propagation. Unlike protons or
nuclei, neutrinos created at high redshifts reach Earth because
they rarely interact and only undergo adiabatic energy losses
and flavor mixing. Figure~\ref{fig-1} (right panel) shows the flux of cosmogenic neutrinos associated
to the best-fit obtained in \cite{Heinze:2015hhp}.
The bottom curve--marked ``TA fit min''--is the envelope
of all possible neutrino fluxes allowed by the cosmic-ray fit. It
is in tension with the TA UHECR data at 99.7\% C.L. Moreover, the event rate
can be calculated using the fluxes resulting from the fit, as for example explained in
\cite{Baerwald:2014zga}: in particular, ``TA fit min'' yields 4.9 events. Since only one
event was observed, this flux can be excluded at 95\% C.L. We have then demonstrated that the limits on neutrino flux can be used as a test of the proton dip model completely independent of composition data.\\
The presence of protons in UHECRs can also be constrained by considering the photons that are produced during propagation, in the light of the recent results obtained by the Fermi-LAT collaboration about the origin of the EBL \cite{Ackermann:2014usa}. These constraints, additional or complementary to the ones given by the expectations from neutrino production during propagation, are discussed in \cite{Berezinsky:2016jys,Supanitsky:2016gke,vanVliet:2016dyx,Kalashev:2016xmy}.

\section{Hypothesis of mixed composition}
\label{sec-2}
\begin{figure*}[b]
\centering
\includegraphics[width=7cm]{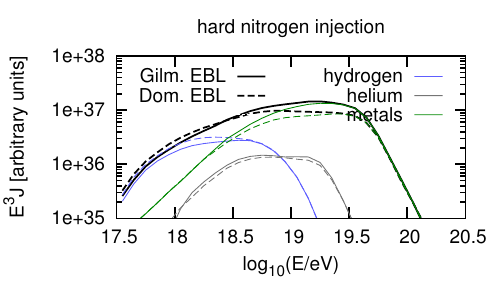}
\includegraphics[width=7cm]{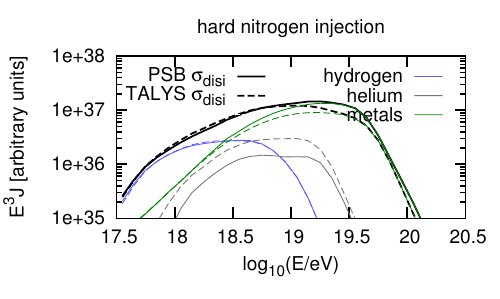}
\caption{Effect of different EBL models (left panel) and of different photo-disintegration models (right panel) on propagated UHECR fluxes \cite{Boncioli:2015pds}.}
\label{fig-4}     
\end{figure*}

\begin{figure}[!htb]
\centering
\begin{tabular}{c c}
\includegraphics*[width=0.55\textwidth]{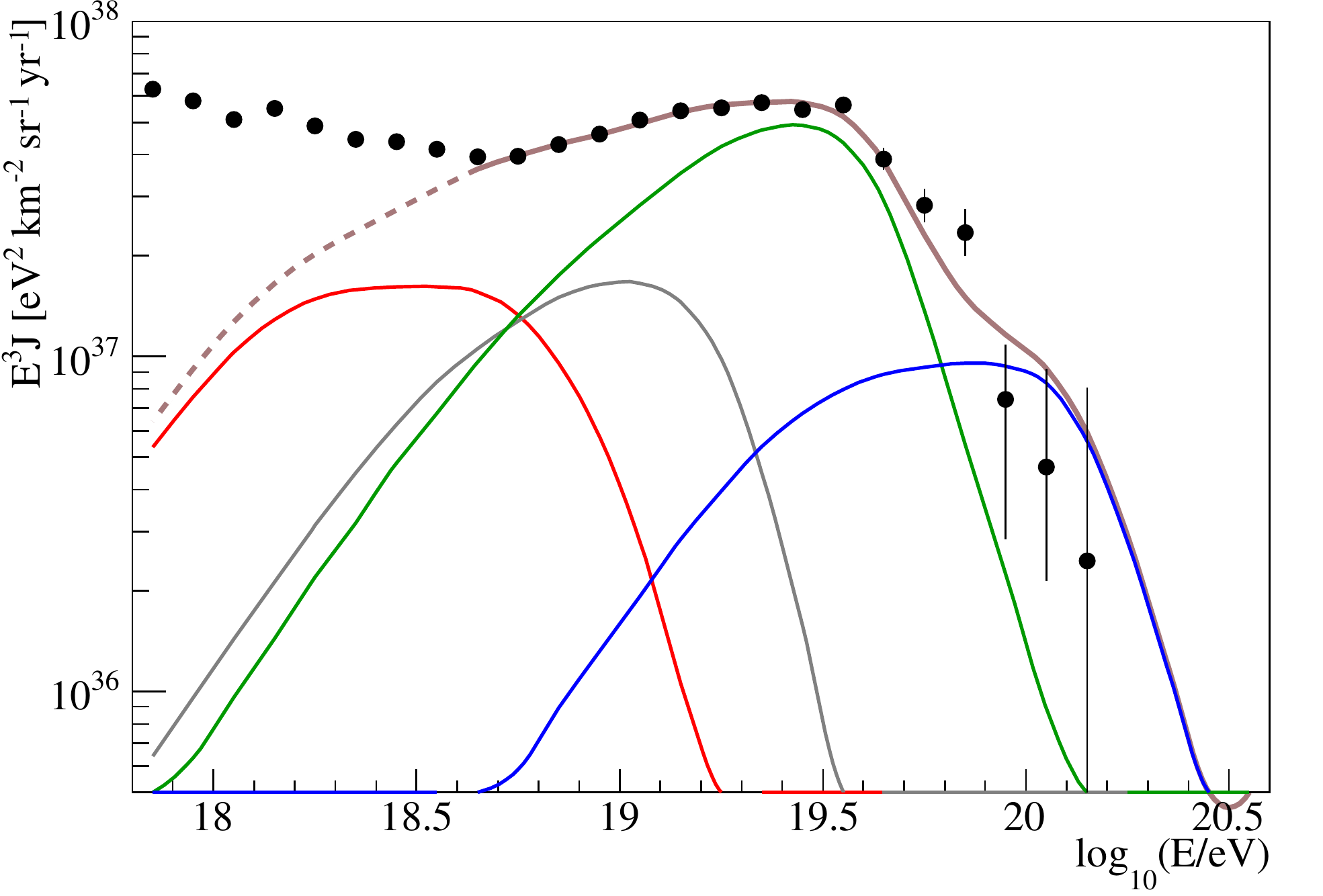}\\
\includegraphics*[width=0.3\textwidth]{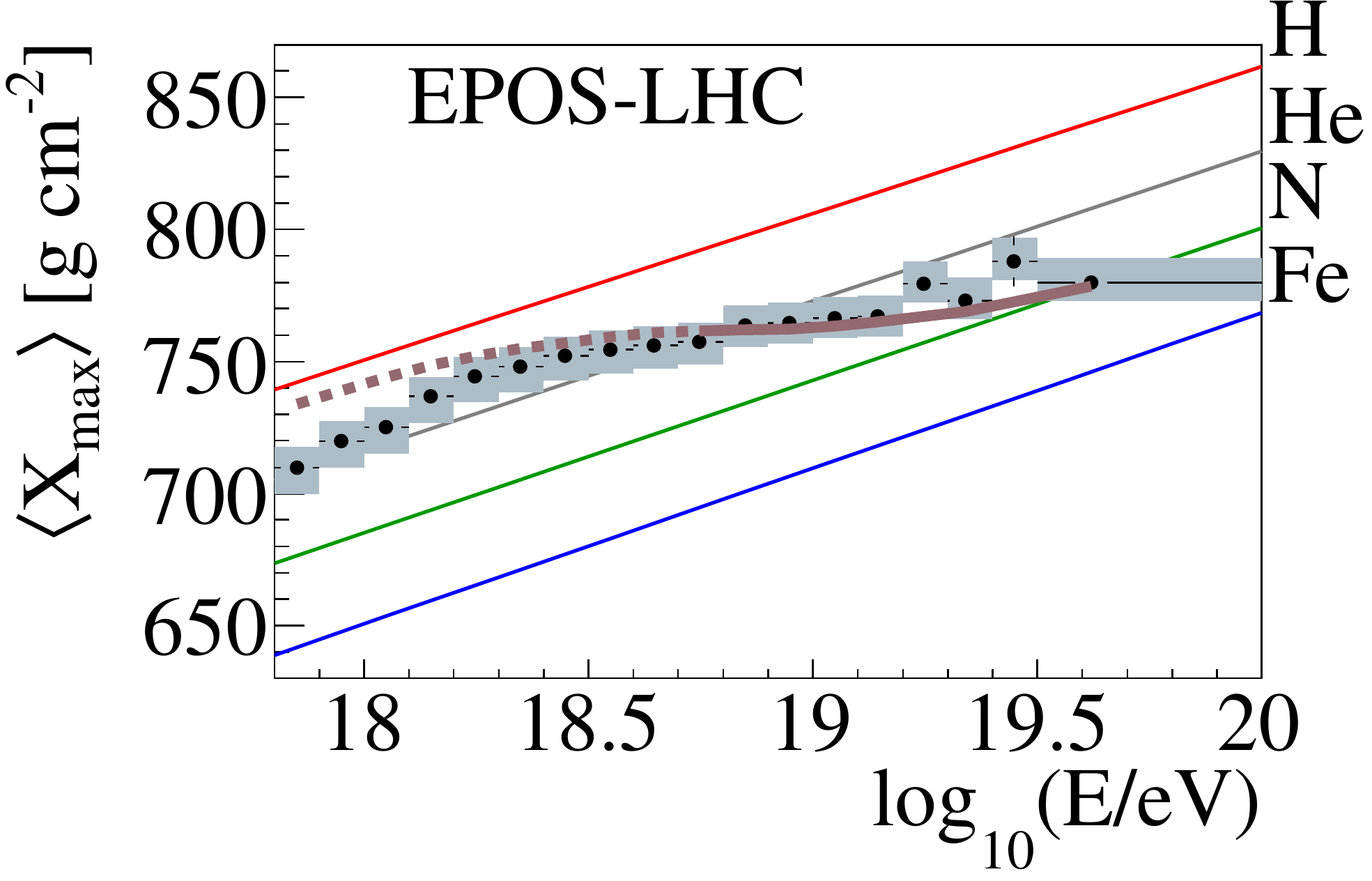} 
\includegraphics*[width=0.3\textwidth]{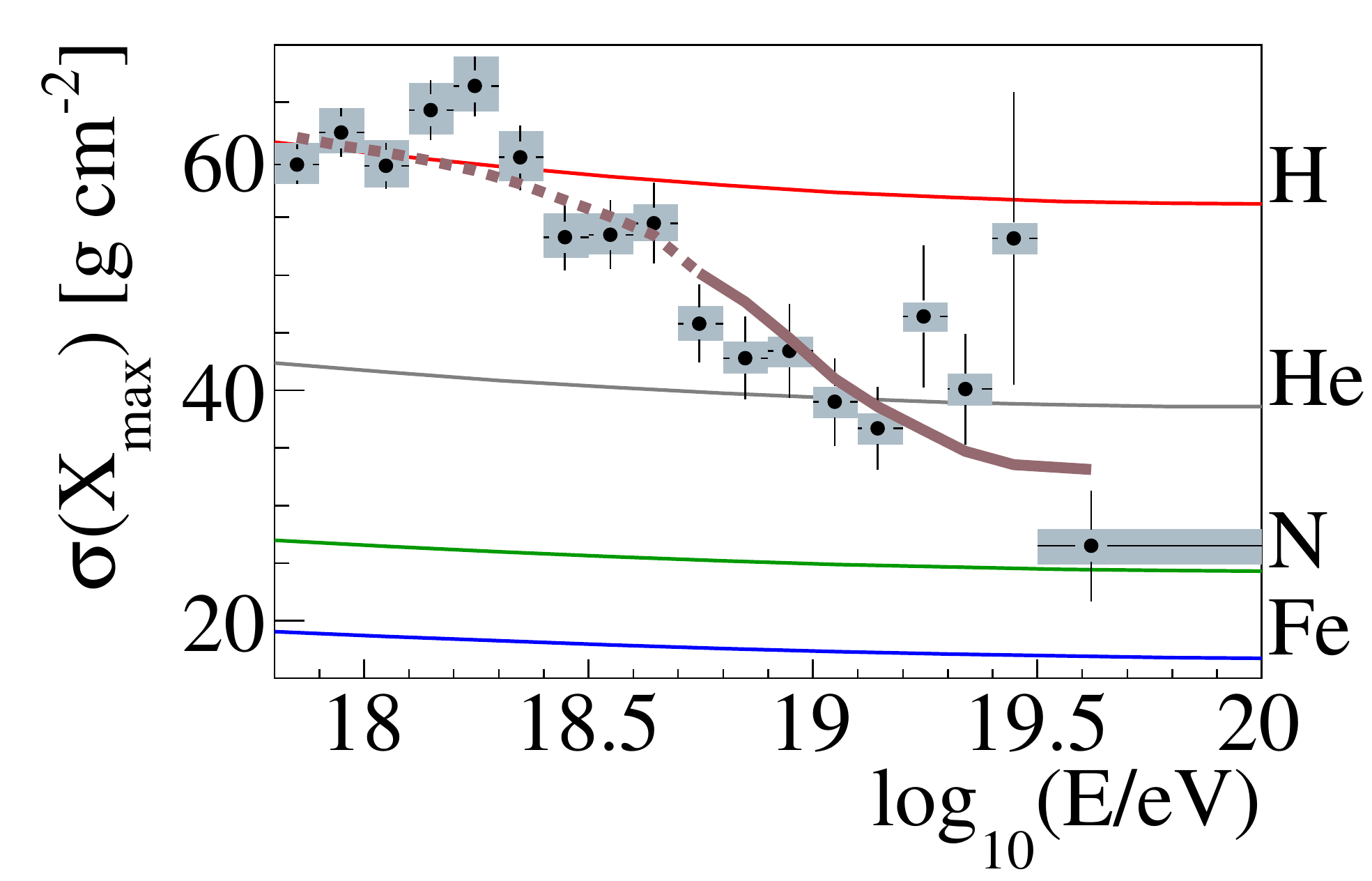}
\end{tabular}
\caption {Simulated UHECR fluxes, $\langle X_{\mathrm{max}} \rangle$ and $\sigma(X_{\mathrm{max}})$ in the best-fit scenario found in \cite{diMatteo:2015} (thick brown: total; red: $A=1$; grey: $2\leq A \leq4$; green: $5\leq A \leq 26$; blue: $A \geq 27$) compared to Auger data \cite{Valino:2015,Porcelli:2015}.}\label{fig-5}
\end{figure}

Several analytical and Monte Carlo codes have been developed to
simulate the propagation of UHECRs. Here we mainly discuss the accuracy of our knowledge of relevant quantities for modeling nuclei interactions, making use of a publicly available Monte Carlo code, {\it SimProp} \cite{Aloisio:2012wj}.
The expansion of the Universe, the spectrum and evolution of the CMB and the cross section for pair production are known with good accuracy and enter in calculating interactions of both
protons and nuclei. For the case of mixed composition, photo-disintegration processes are possible both with CMB and EBL photons. The quantities here involved are poorly known.\\
The choice of the EBL model does not significantly affect the UHECR spectrum at Earth if the composition is assumed as pure protons. It is well known that it is important for neutrino production (however, this uncertainty is smaller than the ones in the cosmological evolution of sources \cite{Aloisio:2015ega}).
\begin{floatingfigure}[p]{.5\textwidth}
\centering
\includegraphics[width=6cm]{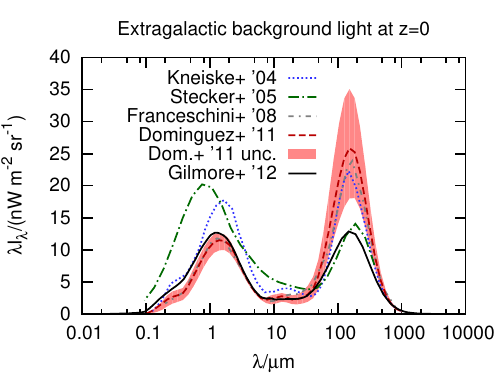}
\caption{Intensity of EBL at $z=0$. See \cite{Batista:2015mea} and references therein.}
\label{fig-2} 
\end{floatingfigure}
In the case of nuclei, due to GDR, at energies $\approx 10^{19}$ eV photo-disintegration can occur via interaction with photons with energies $\approx$ 10 $-$ 100 meV in the laboratory frame: the EBL plays a fundamental role in the propagation of nuclei.
As shown in Figure~\ref{fig-2}, the discrepancies among different models are larger for larger wavelengths, which are the most relevant ones for UHECR propagation. The effect of these uncertainites is shown in Figure~\ref{fig-4} (left panel): the stronger far-IR peak in the EBL spectrum in Dominguez EBL \cite{Dominguez:2010bv} with respect to the Gilmore EBL model \cite{Gilmore:2011ks} causes more interactions, resulting in a higher production of light particles at Earth and a softer spectrum.\\
The other main source of uncertainties in calculations of cosmic rays interactions comes from nuclear physics. Data on photo-nuclear cross sections are scarse and the existent models use photo-neutron cross sections where available, or empirical parametrizations \cite{Boncioli:2016lkt}.
In Figure~\ref{fig-3} the experimental situation on photo-nuclear cross sections is shown, based on data from EXFOR database \cite{Otuka2014272} in the GDR and quasi-deuteron energy band, for nuclear isotopes interesting for cosmic-ray astrophysics.
The availability of nuclear models is also shown; moreover, it can be seen that there are models, like TALYS \cite{Koning:2007}, which predict the population of isotopes off the main diagonal, while more simplified models as PSB \cite{Puget:1976nz,Stecker:1998ib} have only one isotope for each $A$.
Moreover, the TALYS model allows the ejection of small fragments and not only nucleons as in the PSB model.
The effect of the use of these different models in UHECR propagation is shown in Figure~\ref{fig-4} (right panel): the spectrum of residual primary nuclei at Earth is expected to be more eroded at the highest energies in the TALYS model than in the PSB one, because of the more efficient photo-disintegration.
\begin{floatingfigure}[p]{.5\textwidth}
\begin{center}
  \centering
  \includegraphics[width=6.5cm]{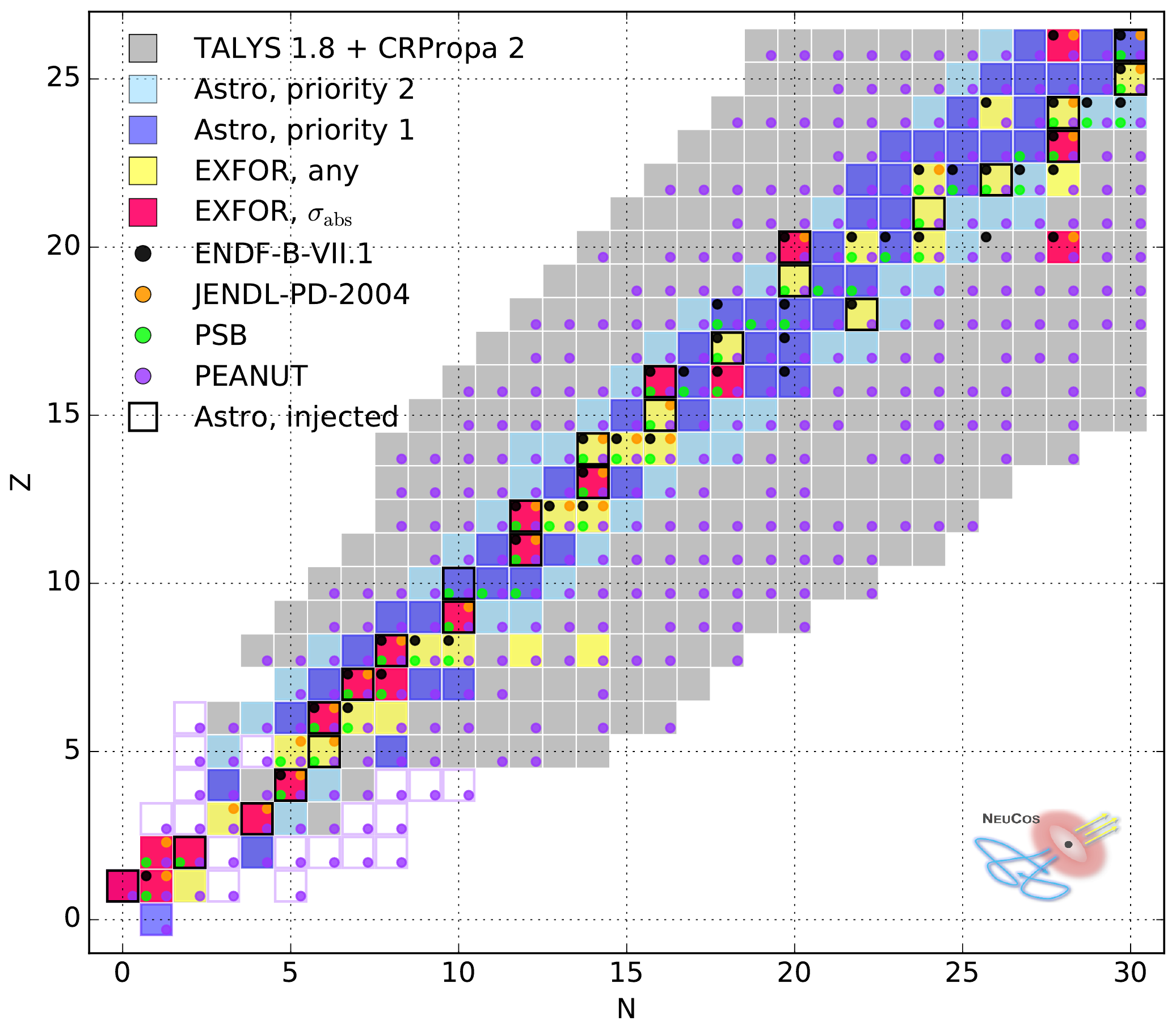}
  \caption{Experimental situation versus astrophysical requirements for nuclear isotopes interesting for cosmic ray astrophysics (gray boxes, from TALYS and CRPropa). Experimental measurements (from EXFOR database \cite{Otuka2014272}) are marked by red and yellow boxes for the total absorption and any inclusive cross sections, respectively. Models are marked by dots (see \cite{Boncioli:2016lkt}, from which the figure is taken, and references therein).}\label{fig-3}
\end{center}
\end{floatingfigure}
For the same reason, the flux of small fragments is enhanced.\\
A fit of the Auger spectrum \cite{Valino:2015} and $X_{\mathrm{max}}$ data \cite{Porcelli:2015} with a simplified model of UHECR sources has been performed \cite{diMatteo:2015,Boncioli:2015pds}, with the assumptions of identical sources homogeneously distributed in a comoving volume. The injection consists only of ${}^{1}$H, ${}^{4}$He, ${}^{14}$N and ${}^{56}$Fe. In this study two different simulation codes for propagation have been used, {\it SimProp} and CRPropa \cite{Batista:2016yrx} with different settings for the EBL and cross section models, in order to quantify the effect of such differences on the fit. In particular, if the propagation is simulated with {\it SimProp} v2r3 \cite{Aloisio:2016tqp} using the Gilmore EBL \cite{Gilmore:2011ks} and the PSB photo-disintegration model \cite{Puget:1976nz,Stecker:1998ib} and air showers are simulated with CONEX using the EPOS-LHC hadronic interaction model, the best fit is found with relatively hard spectral index, relatively low cutoff rigidity and a very metal-rich composition.
The corresponding simulated spectrum and average and standard deviation of $X_{\mathrm{max}}$ distributions are shown and compared to Auger data in Figure \ref{fig-5}. The fit has been repeated by using different MonteCarlo codes, EBL models, photo-disintegration models, air interaction models and with data shifted by their systematical uncertainties, finding that the best-fit parameter values are strongly dependent on the models used, though certain features of the best fit are shared by almost all scenarios.
\section{Conclusions}
\label{concl}
At the present time, the proton dip, ankle and mixed-composition models all remain valid. The predicted flux of cosmogenic neutrinos can be used as a tool to test the proton dip model, as done in \cite{Heinze:2015hhp}, independently from composition data.
This method has in principle the power to rule out candidates to UHECR acceleration that have very high source evolution. However, predictions of neutrinos from interactions inside the source and their connections with the cosmic-ray escape mechanism can also be used as a complementary information to the cosmogenic neutrino studies \cite{Baerwald:2013pu}.\\
One possible alternative to the proton dip model is the mixed-composition scenario, that has been tested for example in \cite{diMatteo:2015} by fitting the Auger spectrum and composition results. In a mixed-composition hypothesis several poorly-known quantities enter in the modeling of extragalactic propagation, such as the EBL models and the photo-disintegration cross sections. The fit of experimental data is affected by uncertainties in the models.\\
Uncertainties in details of photon field and photo-disintegration affect not only the extragalactic propagation but also the modeling of interactions in the source ambient. The missing information from the nuclear physics side is described in \cite{Boncioli:2016lkt} and it is valid for cosmic-ray astrophysics in general.

\section*{Acknowledgments}
This work has received funding from the European Research Council (ERC) 
under the European Union's Horizon 2020 research
and innovation programme (Grant No. 646623).


%
%
%

\end{document}